\documentstyle[12pt,A4]{article}
\begin{document}
\def\refscript#1{\raisebox{0.9ex}{\scriptsize #1)}}
\renewcommand{\thefootnote}{\fnsymbol{footnote}}
\begin{flushright}
Kanazawa-96-06\\
March, 1996
\end{flushright}

\begin{center}
{\large $SO(10)$ Models on $Z_6$\ Orbifold with Dual Wilson Line} \bigskip\\
\vspace{1ex}
\end{center}
\vspace{2.0cm}
\begin{center}
Noriyasu OHTSUBO\footnote[1]{e-mail:ohtsubo@hep.s.kanazawa-u.ac.jp} and 
Masafumi SHIMOJO\footnote[7]
{e-mail:shimo0@ei-phi.ei.fukui-nct.ac.jp}
\bigskip \\
{\sl $^{*}$Department of Physics} \\
{\sl Kanazawa Institute of Technology} \\
{\sl Ohgigaoka 7-1, Nonoichi, Isikawa 921}
\medskip \\
{\sl $^{**}$Department of Electronics and Information Engineering} \\
{\sl Fukui National College of Technology} \\
{\sl Sabae, Fukui 916}
\end{center}
\vspace{1.0cm}
\setlength{\baselineskip}{26pt}
\centerline{ABSTRACT}
'Dual' is a promising key word in the particle physics at present.
The string theory is dual in any sense.
The observed sector and the hidden sector are dual on the 10-dim. 
$E_8\times E_8$ heterotic string. 
We find $Z_6$ orbifold models preserving the duality under a torus 
compactification and realizing $SO(10)$ SUSY GUT in the obserbed sector 
under a twist division.
\newpage

Duality appears in various phases of string theory. As for the gauge group 
of the ten dimensional heterotic string theory\refscript{\ref{gross}}, the 
weight lattices of gauge degrees of freedom of $E_8$(observable) sector and
$E_8^{\prime}$(hidden) sector are dual so that it is natural for a four 
dimensional string theory with this duality to result from the theory.

It is possible to preserve the duality in the toroidal compactification
\refscript{\ref{Narain}} which makes use of $Z_6$ translation group to 
reduce the space-time dimensions. When we represent the gauge degrees of 
freedom as $E_8\times E_8^{\prime}$\ root momenta $P^I(I=1...16)$ of the 
left-moving bosonic variables, the representation of the translation group 
in the gauge degrees of freedom is so-called a Wilson line $a^I$ and 
physical fields must satisfy the $Z_6$\ singlet condition 
$\sum_{}^{}P^Ia^I=0$\ mod 1.  Such a dual Wilson line as 
$a=({\bf a},-{\bf a})$ where ${\bf a}$\ is an 8-d shift vector on top of 
the $E_8$\ lattice brings about same gauge group with mutually dual weight 
lattice to the two sectors. 

The toroidal compactification, however, leads to four dimensional 
string theories with gauge groups of rank 22 and {\it N}=4 supergravity
\refscript{\ref{Narain}}. 
In order to break the large symmetry, we can follow the 
orbifold compactification. The scheme makes use of space group S whose 
elements consist not only of a translation v but also of a twist 
$\theta$ which is an element of point group. 
For the bosonized NSR fermions, the twist $\theta$ is expressed by 
${\rm e}^{2\pi i\xi}$\ and {\it N}=1 supersymmetry requires that $\xi$ is 
given by 
\begin{equation}
\xi=\frac{1}{{\rm N}}(1,m,-1-m,0),
\label{xi}
\end{equation} 
where $m$\ is an integer and N is the order of orbifold restricted to one of 
integers\\
{\large 3,\ 4,\ 6,\ 7,\ 8,\ 12}\ \refscript{\ref{ohtsubo}}. 

In fact, the twist breaks down also the duality and 
it is impossible to find out a dual embedding from 
the point group into $E_8\times E_8^{\prime}$ gauge degrees of freedom. The 
reason is as follows: we embed both point group and translation 
group of S to the Cartan subalgebra of $E_8\times E_8^{\prime}$ and 
express the representation as a shift vector $V_f^I=v^I+a^I(I=1..16)$ 
on the $E_8\times E_8^{\prime}$\ weight, 
where $v^I$ is a shift corresponding to a twist $\theta^1$\ and $a^I$\ 
is a Wilson line. We assume a dual shift 
$V_f=\frac{1}{2{\rm N}}({\bf x},-{\bf x})$, 
where ${\bf x}$\ is an 8-d vector on the $E_8$ weight lattice. 
Since $\frac{{\bf x}}{2}$ is a weight of $E_8$, $x^I(I=1..8)$\ are all 
even or all odd integers in an orthonormal base and ${\bf x}^2$ must be 
a multiple of eight. Also modular invariance condition of the orbifold  
partition function gives a restriction 
\begin{equation}
{\rm N}(V_f^2-\xi^2)=2k,
\label{modinv}
\end{equation}
where $k$\ is an integer.
Substituting Eq.(\ref{xi}) to the Eq.(\ref{modinv}), we obtain that
\begin{equation}
\sum_{I=1}^{8}(x^I)^2=4(k{\rm N}+1)+4m(m+1).
\label{sumx}\end{equation}
If the order N was even, the r.h.s. was not a multiple of eight 
and the modular invariance was inconsistent with the duality.

On $Z_3$\ and $Z_7$\ orbifolds, we have looked for dual shifts which give 
non-abelian gauge groups of GUT models and MSSMs, i.e., $SU(5),SO(10)$\ and 
$SU(3)\times SU(2)$. But no dual $V_f^I$\ with dual Wilson lines which gives 
such groups exists.

It is possible, whether the order N is primary or not, 
to map neither an overall element of S nor a twist to a dual shift through a 
modular invariant embedding, because the modular invariance condition (
\ref{modinv}) for a shift $V_f^I$ applies also to the shift $v^I$. Here if 
we respect the duality, it is necessary to insist that space-time of 
extra dimensions is not directly compactified to a orbifold but to a 
torus $T^6$ with a dual Wilson line at first.

To verify the possibility, we investigate $SO(10)$\ GUT models with 
a dual Wilson line on $Z_6$ orbifold. We choose the $SO(10)$\ simple 
roots $R^I$\ as 
\begin{equation}\begin{array}{l}
(0^4,1,0,-1,0;0^8),\ (0^6,1,-1;0^8),\ (0^5,1,0,1;0^8), \\
(0^3,1,0,-1,0^2;0^8),\ \frac{1}{2}(1,-1,-1,-1,-1,-1,-1,1;0^8),
\end{array}
\label{root}
\end{equation}
and look for combinations of two shifts $V_{f_1}^I$ and $V_{f_2}^I$ 
which obey modular invariant condition (\ref{modinv}) and $SO(10)$\ 
invariant conditions that $R\cdot V_{f_i}=0\ {\rm mod}\ 1\ (i=1,2)$ and that 
$R^{\prime}\cdot V_{f_1}\neq 0\ {\rm mod}\ 1$ or 
$R^{\prime}\cdot V_{f_2}\neq 0\ {\rm mod}\ 1$ for roots $R^{\prime}$\ 
other than $SO(10)$. When we take $V_{f_1}^I$ as a shift $v^I$ 
corresponding to a twist, the Wilson line is given by 
$a^I=V_{f_2}^I-V_{f_1}^I$. We pick out combinations with a dual Wilson 
line and examine massless matter fields given by the shifts. 
The detail of massless conditions and $Z_6$ singlet condition that 
physical fields must satisfy is found in Ref.\ref{ohtsubo}),\ref{sashi}).

The $SO(10)$\ of the roots (\ref{root}) contains flipped $SU(5)\times 
U(1)_X\supset SU(3)_c\times SU(2)_L\times U(1)_Y$\ so that axial anomaly 
related to weak hypercharge $Y$\ of matter fields automatically vanishes. 
We can assign {\bf 16} of the $SO(10)$ to quark, lepton fields 
and assign {\bf 10} to two Higgs doublets, respectively\refscript{\ref{sashi}}.

We choose spectra which contain just three generations of {\bf 16} 
except for pairs of {\bf 16} and ${\bf 16}^*$. We exclude models with chiral 
anomaly with respect to the gauge group of the hidden sector and so-called 
T-duality anomaly\refscript{\ref{lust}}. Up to now, we have obtained three 
models on $Z_6$-I orbifold. For one of the models, we give the shift $v^I$, 
the Wilson $a^I$ and the spectrum of matter fields to the Table I, 
where $\theta^k$ represents the $k$-twisted sector. The gauge group of the 
hidden sector is $SO(10)\times SU(2)$. \\ 
\\
({\bf Table I})\\

When we suppose a model on torus with the dual Wilson line in the table, 
we get {\large $N=4$ dual $E_6\times SU(3)$ model} with matter fields of 
the adjoint representation in both observable and hidden sectors.

Also, the other two models are given by the 
Wilson line of the table, while the shifts which lead to the 
models are
$v=\frac{1}{12}(-3,-3,-1,1,-1,1,-1,-1;3,-3,5,1,-1,1,-1,-1)$,  
$v=\frac{1}{12}(-3,-3,-1,1,-1,1,-1,-1;2,2,-2,2,0,0,-2,-2)$. The gauge 
groups of the models are  $SO(10)\times SO(10)^{\prime}\times SU(2)^{\prime}$ 
and $SO(10)\times SU(5)^{\prime}\times (SU(2)^{\prime})^2$. 

{\bf References}

\begin{enumerate}
\item D.J. Gross, J.A. Harvey, E.Martinec and R.Rohm, Phys. Rev. Lett. 
54(1985)502; Nucl. Phys. B256(1985)253; B267(1986)75. 
\label{gross}
\item K.S. Narain, M.H. Sarmadi, E. Witten, Nucl. Phys. B279(1987)369 
\label{Narain}.
\item T. Kobayashi and N. Ohtsubo, Int. J. Mod. Phys. A. Vol 9, No1(1994)87. 
\label{ohtsubo}
\item Hikaru Sato, M.Shimojo, Phys. Rev.D48(1993)5798. \label{sashi}
\item L.E. Ib\`a\~nez, D. L\"ust, Nucl. Phys. B382(1992)305. \label{lust}
\end{enumerate}
\newpage
\begin{flushright}
N.Ohtsubo and M.Shimojo\\
\end{flushright}
\noindent
\setcounter{page}{1}
\begin{table}[h]
Table I. Mass spectrum of a $SO(10)\times SO(10)^{\prime}\times 
SU(2)^{\prime}$ model through\\
$v=\frac{1}{12}(-3,-3,-1,1,-1,1,-1,-1;6,6,2,-2,2,-2,2,2)$ and \\
$a=\frac{1}{6}(3,1,3,-1,1,-1,1,1;-3,-1-3,1,-1,1,-1,-1)$. 
\vspace{0.5cm}\\
\begin{tabular}{|l|l|l|l|}
\hline 
\ \ \ $\theta^0$ & \ \ \ $\theta^1$ & \ \ \ $\theta^2$ & \ \ \ $\theta^3$ \\
\hline
2(16;1,1)& (10;1,1) & 4(10;1,1) & 6(16;1,1),5$(16^*;1,1)$,11(10;1,1) \\
(1;16,1) & $(1;16^*,1)$,2(1;10,1)   & 4(1;10,1) & \\
(1;1,2)  & 6(1;1,2)                &  21(1;1,2) & \\
3(1;1,1)  & 24(1;1,1) & 41(1;1,1) & 61(1;1,1) \\ 
\hline
\end{tabular}
\end{table}
\end{document}